\pdfoutput=1

\documentclass[11pt]{article}

\usepackage[final]{acl}

\usepackage{times}
\usepackage{latexsym}

\usepackage[T1]{fontenc}

\usepackage[utf8]{inputenc}

\usepackage{microtype}

\usepackage{inconsolata}

\usepackage{calc}
\usepackage{amsmath}
\usepackage{caption}
\usepackage{multirow}
\usepackage{amsfonts}
\usepackage[font=footnotesize]{subcaption}
\usepackage{booktabs}
\usepackage{pdfpages}
\usepackage{xcolor}
\usepackage{tabularx}
\usepackage{bbm}
\usepackage{graphicx}
\usepackage{algorithm}
\usepackage{algorithmic}
\usepackage{enumitem}

\newcommand{\horns}{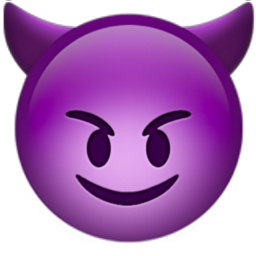}

\newcommand{\sparkles}{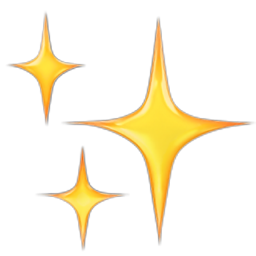}

\newcommand{\pleadingface}{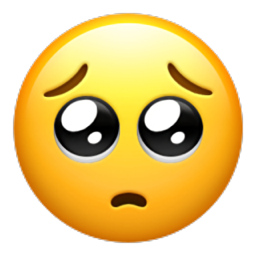}

\newcommand{\fire}{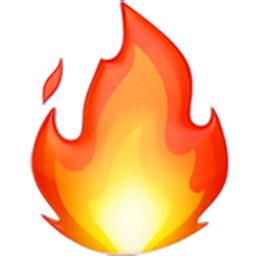}

\newcommand{\hearteyes}{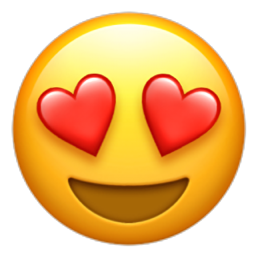}
\newcommand{\eyes}{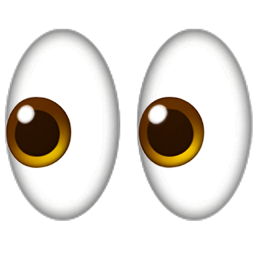}

\newcommand{\whiteheart}{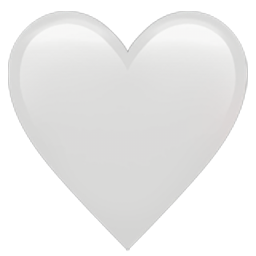}

\newcommand{\pensiveface}{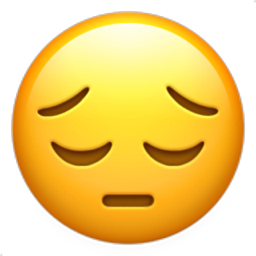}

\newcommand{\thumbsup}{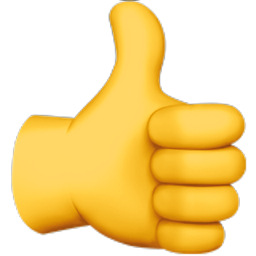}

\newcommand{\partypopper}{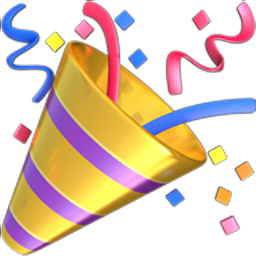}

\newcommand{\thinkingface}{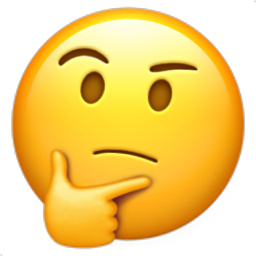}

\newcommand{\bigeyes}{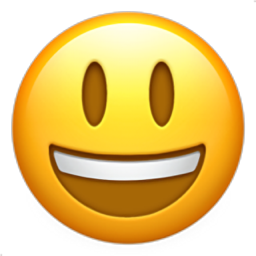}

\newcommand{\smilingfacewithhearts}{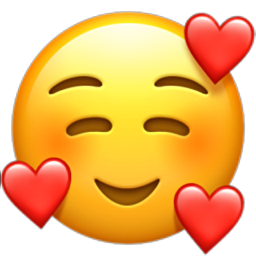}
\newcommand{\smilingfacewithsmilingeyes}{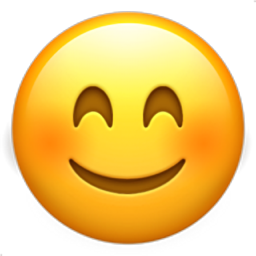}

\newcommand{\ninja}{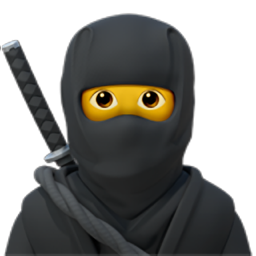}
\newcommand{\smilingfacewithtear}{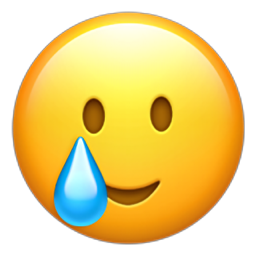}
\newcommand{\magicwand}{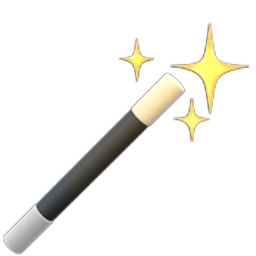}
\newcommand{\pinchedfingers}{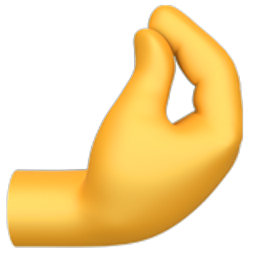}
\newcommand{\coinemoji}{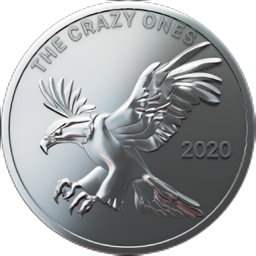}
\newcommand{\peoplehugging}{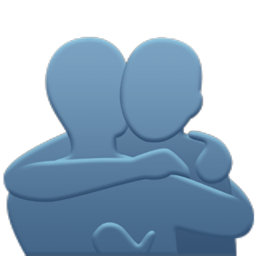}
\newcommand{\wearyface}{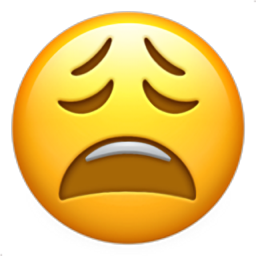}
\newcommand{\collision}{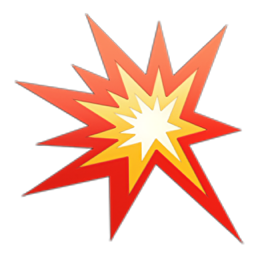}
\newcommand{\hundredpoints}{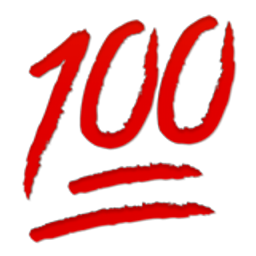}
\newcommand{\blueheart}{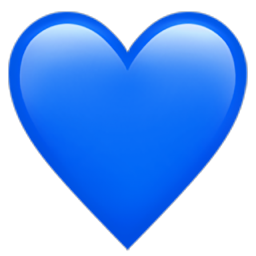}
\newcommand{\disguisedface}{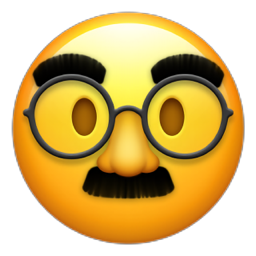}
\newcommand{\ladder}{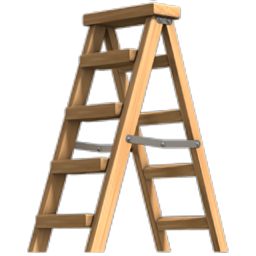}
\newcommand{\blueberries}{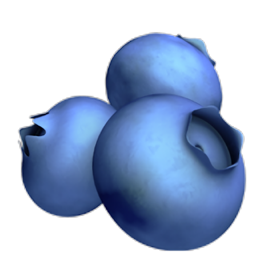}
\newcommand{\facewithspiraleyes}{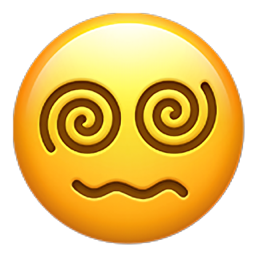}
\newcommand{\flushedface}{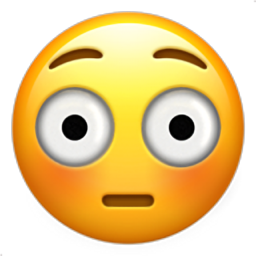}
\newcommand{\mendingheart}{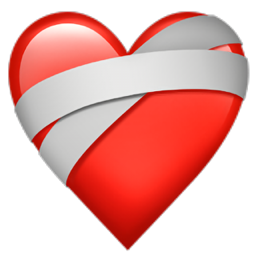}
\newcommand{\droplets}{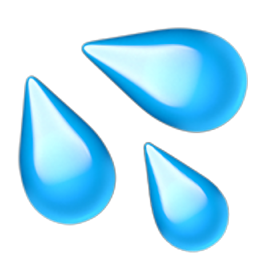}
\newcommand{\woozyface}{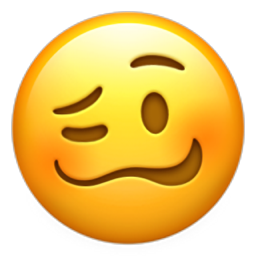}
\newcommand{\faceexhaling}{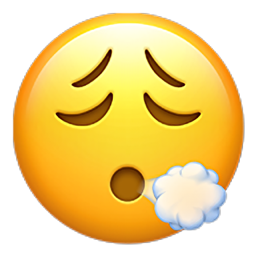}
\newcommand{\faceinclouds}{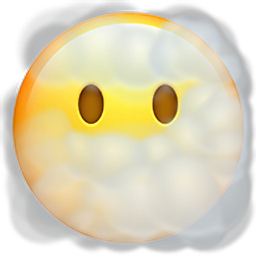}
\newcommand{\hotface}{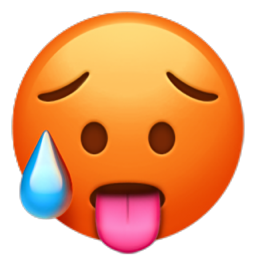}
\newcommand{\partyingface}{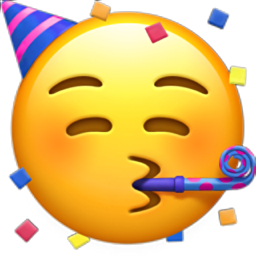}
\newcommand{\coldface}{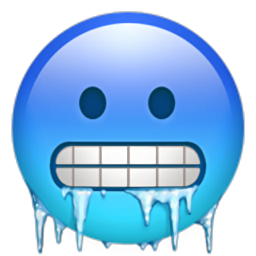}
\newcommand{\teddybear}{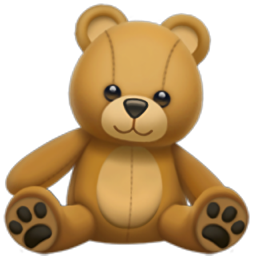}
\newcommand{\firecracker}{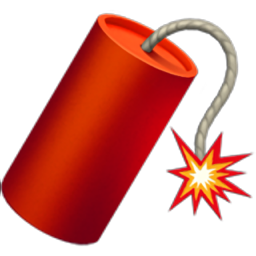}
\newcommand{\cupcake}{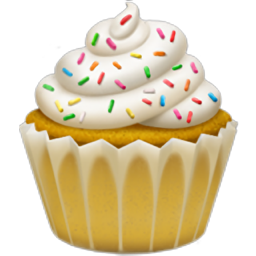}
\newcommand{\foot}{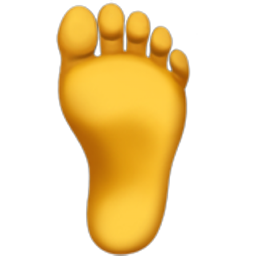}
\newcommand{\personstanding}{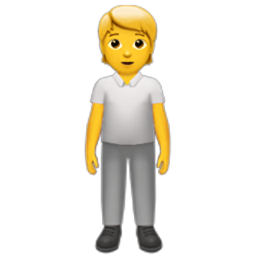}
\newcommand{\yawningface}{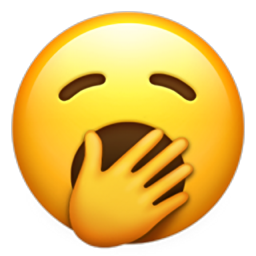}
\newcommand{\brownheart}{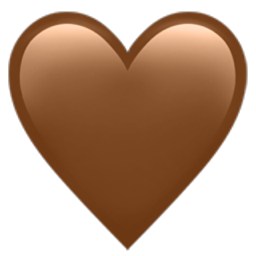}
\newcommand{\pinchinghand}{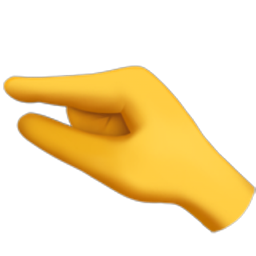}
\newcommand{\dropofblood}{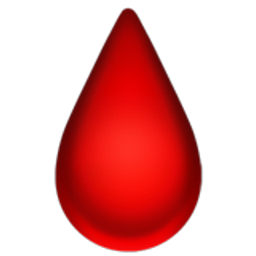}
\newcommand{\ringedplanet}{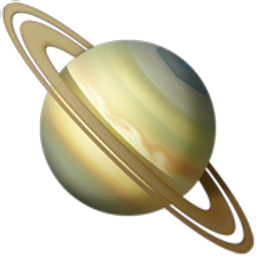}
\newcommand{\whitecane}{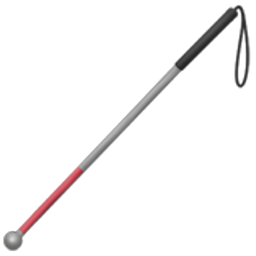}
\newcommand{\purplecircle}{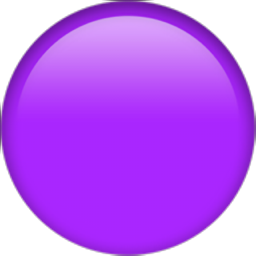}
\newcommand{\otter}{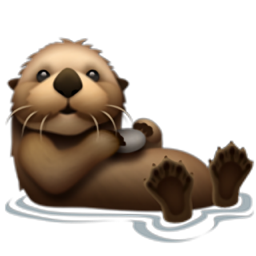}
\newcommand{\pottedplant}{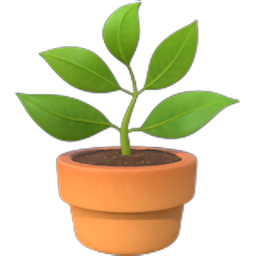}
\newcommand{\flatbread}{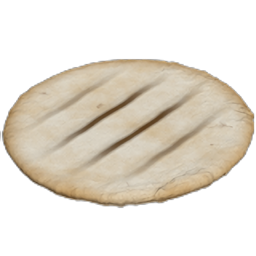}
\newcommand{\placard}{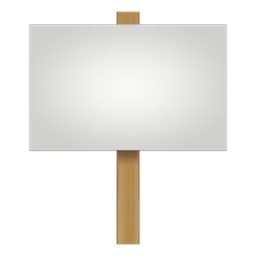}

\usepackage[export]{adjustbox}
\newcommand{\MyEmoji}[1]{\includegraphics[width=1em,valign=t]{#1}}

\title{From Adoption to Adaption: Tracing the Diffusion of New Emojis on Twitter}

\author{Yuhang Zhou \\
    University of Maryland \\
    College Park, USA \\
  \texttt{tonyzhou@umd.edu} \\\And
  Xuan Lu \\
  University of Arizona \\
  Tucson, USA \\
  \texttt{luxuan@arizona.edu} \\\And
  Wei Ai \\
  University of Maryland \\
    College Park, USA \\
  \texttt{aiwei@umd.edu} \\}

\begin{document}
\maketitle
\begin{abstract}
In the rapidly evolving landscape of social media, the introduction of new emojis in Unicode release versions presents a structured opportunity to explore digital language evolution. Analyzing a large dataset of sampled English tweets, we examine how newly released emojis gain traction and evolve in meaning. We find that community size of early adopters and emoji semantics are crucial in determining their popularity. Certain emojis experienced notable shifts in the meanings and sentiment associations during the diffusion process. Additionally, we propose a novel framework utilizing language models to extract words and pre-existing emojis with semantically similar contexts, which enhances interpretation of new emojis. The  framework demonstrates its effectiveness in improving sentiment classification performance by substituting unknown new emojis with familiar ones. This study offers a new perspective in understanding how new language units are adopted, adapted, and integrated into the fabric of online communication.
\end{abstract}

\section{Introduction}
\label{sec:intro}

The language landscape of the Web era is ever-evolving, characterized by the emergence and evolution of new language units. Individuals have creatively crafted out-of-vocabulary language units, such as Internet memes and viral hashtags, to encapsulate and convey complex ideas, sentiments, and cultural phenomena, fostering shared lexicons that resonate across digital communities. Understanding the adoption and adaptation of these language units is crucial for gaining insights into the dynamic nature of online communication, the information diffusion process in social networks, and the underlying social trends and movements. However, analyzing the dynamics of these emerging language units in online communication presents unique challenges. These units lack universal conventions and standards and their characteristics may vary during diffusion, making it complex to track their initial appearances, early adoption, and frequency of use.

As a recent addition to this landscape, emojis offer a distinctive opportunity to explore the diffusion and evolution of new language units. Emojis are visual symbols that are embedded into text. These non-verbal symbols go beyond a single word or phrase, encapsulating rich semantics spanning a wide spectrum of emotions, actions, objects, and concepts. Originating as emoticons in the early Internet culture, emojis have evolved into a standardized and universally recognized visual language. Unlike other language units like hashtags or internet memes, emojis undergo a standardized process prior to their inclusion in the language. They are proposed to the Unicode consortium, formally defined by the Unicode standard, uniquely coded as Unicode strings, such as U+1F603 for emoji \MyEmoji{\bigeyes}, and then rendered by various platforms. 

Since the Unicode started to adopt emojis in 2010, we have witnessed emojis' remarkable rise on the Web, with their adoption consistently increasing across multiple platforms \cite{rong2022empirical, lu2018first, kejriwal2021empirical, halverson2023content}. New emojis continue to be introduced in response to user requests. From 2018 to 2022, five new versions of emojis (Unicode 11.0 to 15.0) have been released, some of which, like the pleading face emoji (\MyEmoji{\pleadingface}) and the partying face emoji (\MyEmoji{\partyingface}), have gained widespread adoption among Twitter users. 

This standardized approach ensures that emojis maintain a stable form throughout their journey within social networks, enabling precise tracking of emoji adoption and diffusion. These attributes - precise definition, standardized implementation, stable form, and accurate release information — underscore the unique suitability of emojis for studying the evolution of language in online networks.

We take the initiative to study the diffusion of emerging language units on social media through the unique perspective of Unicode-versioned emojis. We investigate the diffusion of newly created emojis introduced in Unicode versions 11.0 to 13.0, across the Twitter platform, particularly the usage frequency and semantic shift as the new emojis cascade through social media. 

Because of the rich semantics, the Unicode definition (such as ``pleading face'') is far from enough to interpret the meaning of an emoji. To this end, we propose an interpretation framework that leverage language models (LMs) to identify words and existing emojis that share similar semantic contexts with the new emojis. Finally, we evaluate the practical implications of our framework by substituting new emojis with semantically similar ones in sentiment classification tasks, which demonstrates the effectiveness of our approach in helping NLP models interpret emerging language units for downstream tasks. 
We summarize our major contributions as follows:
\begin{itemize}[leftmargin=*]
    \item We explore the pattern of emoji diffusion initial adoption to widespread usage, with a focus on frequency and sentiment aspects.
    \item We introduce an interpretation framework to interpret the semantics of new emojis by exploring the words or old emojis with similar semantics.
    \item To validate the effectiveness of our interpretation framework, we replace emojis in texts with surrogates and improve the model performance in the sentiment classification task. 
\end{itemize}

\section{Related Work}
\label{sec:related}

Our work is based on two lines of existing work: the emoji understanding as well as its applications and the information cascade in the social media.

\subsection{Emoji Understanding and Applications}
The prevalence of emojis on many platforms, especially the social media platform, has gained increasing interest from researchers in various areas. They have studied the emoji functions in multiple aspects, such as conveying sentiments \cite{ai2017untangling}, highlighting topics \cite{lu2016emojiusgae}, indicating identities \cite{ge2019identity}, promoting communication \cite{zhou2023emoji}. In addition to \citet{lu2018first} which describes the emoji development process on the GitHub platform and \citet{feng2020new} which examines patterns of new emoji requests, the diffusion and interpretation of emojis in the recent version have not been fully studied. 

Note that many researchers have utilized emojis to help with downstream tasks. Sentiment classification is the primary application to include emoji information \cite{chen2019emojipowered, Felbo_2017, lou2020sentiment, Chen_YuXiao2018, eisner2016emoji2vec}, and later research uses emoji usage to predict the developer dropout \cite{lu2022emojis}, the hashtag used \cite{zhou2022emoji}, or user gender \cite{chen2018gender}. However, these research works do not enable the generalizability of the framework on the newly appeared emojis. In this paper, we propose an emoji substitution method to improve the models' effectiveness on the new emojis without further parameter updates.

\subsection{Innovation Diffusion on the Social Media}

Previous work has noticed and explored the innovation sharing pattern on the social media, such as hashtags, meme, blog articles, and news diffusion \cite{ma2014understanding, johann2019one, spitzberg2014toward, kumpel2015news, yang2012whatyoutag, ahmed2013peek, bakshy2011everyone, cheng2017antecedents, cunha2011analyzing} and proposed multiple models to understand and simulate the information cascade process \cite{zhou2021survey}. 
In addition to content innovation, users create new language units and distribute them through social networks \cite{grieve2018mapping, kershaw2018language} but few works systematically studied the reason, pattern, and interpretation of new language units.

In our work, we explore the diffusion of new emojis to reveal the diffusion mechanism of social media and frame the new emoji cascade as the innovation diffusion process. The previous study explores the factors influencing innovation popularity by adopting the theory of innovation diffusion \cite{rogers2014diffusion, ma2014understanding, steffes2009social, chu2011determinants}. 


\begin{figure*}[t!]
\centering
    \begin{subfigure}{0.32\linewidth}
    \centering
    \includegraphics[width=\columnwidth]{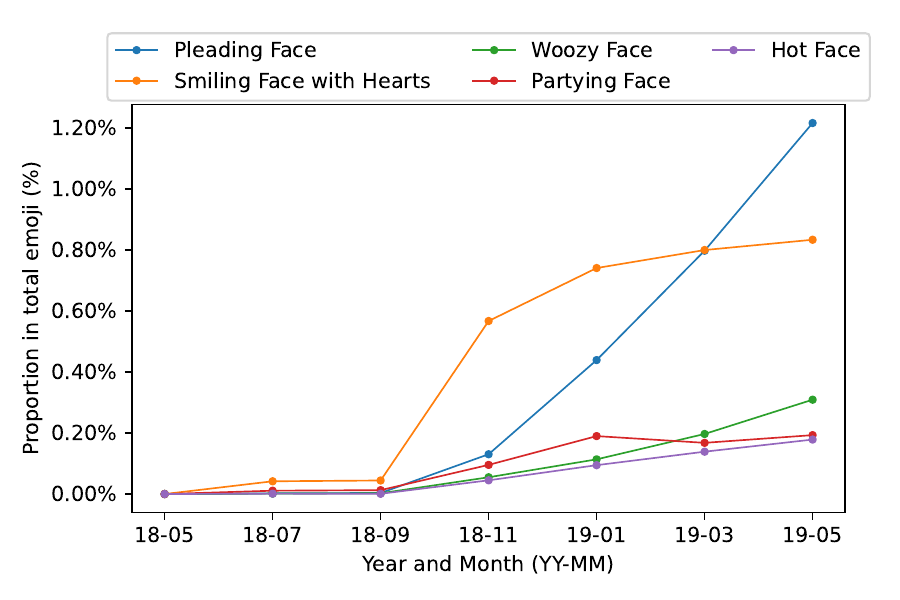}
    \caption{Emoji 11.0}
    \label{fig:emoji11_change}
    \end{subfigure}
    ~ 
    \begin{subfigure}{0.32\linewidth}
    \centering
    \includegraphics[width=\columnwidth]{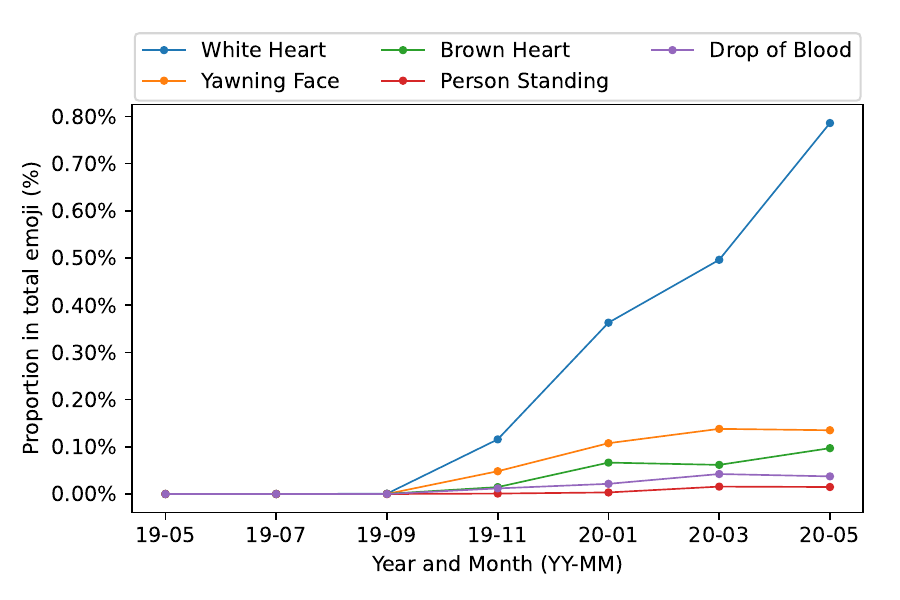}
    \caption{Emoji 12.0}
    \label{fig:emoji12_change}
    \end{subfigure}
    ~
    \begin{subfigure}{0.32\linewidth}
    \centering~\includegraphics[width=\columnwidth]{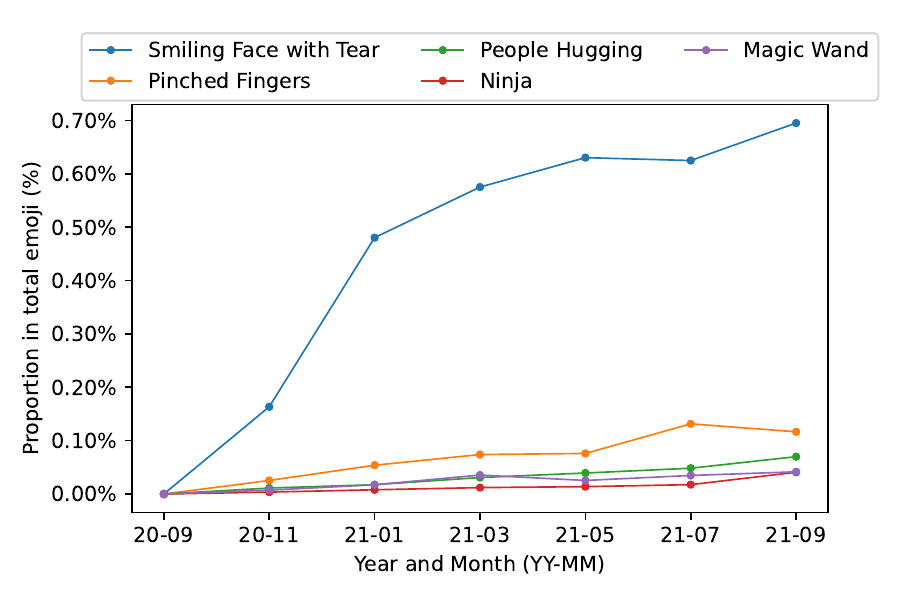}
    \caption{Emoji 13.0}
    \label{fig:emoji13_change}
    \end{subfigure}

\vspace{-0.6em}
\caption{Frequency trends of the top 5 popular emojis in each emoji version over the two years following their first appearance. We show the frequency of emojis every two months, and the y-axis represents the proportion of each emoji in the total of emojis in that month.}
\vspace{-1em}
\label{fig:emoji_frequency_change_month}
\end{figure*}

\section{Understanding Emojis' Diffusion}

We start exploring the diffusion of new emojis by analyzing the their usage frequency and semantic shift. We collected English Tweets from May 2018 to May 2022 using the Twitter API,\footnote{\url{https://developer.twitter.com/en/docs/twitter-api}} during which new emojis are released with Unicode 11.0 (February 2018), 12.0 (March 2019), 13.0 (January 2020). Note that it usually takes a few months for platforms to support inputting and rendering new emojis. For example, the emojis in Emoji 11.0 first appear on Twitter around July 2018. 

\subsection{Emoji Popularity Increases}

We visualize the frequency change of popular emojis of each version in Figure \ref{fig:emoji_frequency_change_month}. Most of their popularity continuously increase over the two-year period following release. However, the adoption rates differs dramatically both between and within emojis. Even for the most popular emojis of the same version, their frequencies may vary by orders of magnitude within months, showing a power-law distribution similar to old emojis \cite{lu2016emojiusgae}. The adoption speed is also uneven over time. For example, the frequency of \MyEmoji{\pleadingface} (pleading face) is still smaller than \MyEmoji{\partyingface} (partying face) and \MyEmoji{\smilingfacewithhearts} (smiling face with hearts) in January 2019. But by May 2019, the popularity of \MyEmoji{\pleadingface} has exceeded all emojis in Emoji 11.0. It suggests that some emojis are diffused to restricted user communities, while others emojis keep spreading and reach larger communities.

\begin{figure}[!htb]
     \centering
     \includegraphics[width=\linewidth]{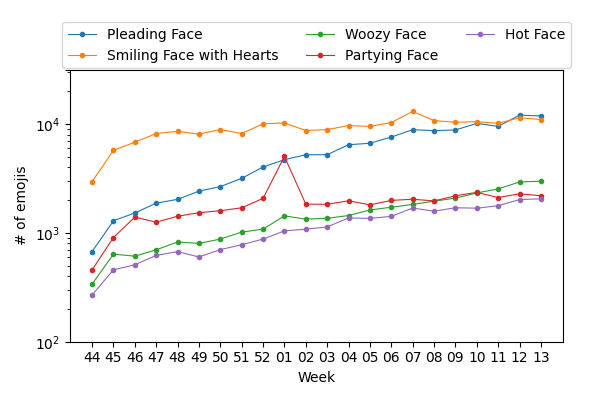}
     \vspace{-2em}
     \caption{Frequency trends by week of the top 5 emojis in Emoji 11.0 from November 2018 to March 2019.}
     \label{fig:emoji_frequency_change_week}
\end{figure}

Seeing the continuous growth of new emojis' popularity after release, we then examine the change in emoji frequency in a more fine-grained time period. We visualize the count of the 5 most popular emojis of Emoji 11.0 at each week from November 2018 to March 2019 in Figure \ref{fig:emoji_frequency_change_week}.
We observe two significant bumps on the lines of \MyEmoji{\partyingface} (partying face) and \MyEmoji{\smilingfacewithhearts} (smiling face with hearts): a bump for \MyEmoji{\partyingface} in Week 1 of 2019 and the a bump for \MyEmoji{\smilingfacewithhearts} in Week 7 of 2019. 

The bumps coincide bursting external events (New Year in Week 1 and Valentine's day in Week 7), which have been discussed in literature as possible triggers for information cascade \cite{zhou2021survey, crane2008robust}. We hypothesize that external events also influence the adoption of new emojis. 
To verify, we examine whether the words associated with the new emoji in that period are related to external events. We collect the tweets with emojis in the first and seventh week of 2019 and calculate pointwise mutual information (PMI) between each emoji $e$ and each word $w$. The PMI equation can be formulated as $\text{PMI}(e, w) = \log \frac{p(e, w)}{p(e)p(w)}$,
where $p(w)$, $p(e)$ and $p(e, w)$ refer to the probability of a tweet containing the word $w$, the emoji $e$, and both of them, respectively. We present the top 10 associated emojis (based on PMI) for emoji \MyEmoji{\partyingface} and \MyEmoji{\smilingfacewithhearts} in different weeks in Table \ref{tab:top_pmi} and highlight the words related to external events, as recognized by the authors.

\begin{table}[!htb]
\centering
\resizebox{\columnwidth}{!}{%
\begin{tabular}{lll}
\toprule
Emoji & Time period     & Top 10 PMI words \\                            \midrule
\MyEmoji{\partyingface}      & W51, 2018 & \begin{tabular}[c]{@{}l@{}}birthday, happy, hope, great, days, \\ we, day, \textbf{year}, amazing, christmas\end{tabular}                  \\ \midrule
\MyEmoji{\partyingface}      & W01, 2019  & \begin{tabular}[c]{@{}l@{}}happy, \textbf{new}, \textbf{year}, birthday, \textbf{2019}, \\ \textbf{happynewyear}, may, everyone, \textbf{years}, \textbf{2018}\end{tabular}             \\ \midrule
\MyEmoji{\smilingfacewithhearts}      & W51, 2018 & \begin{tabular}[c]{@{}l@{}}amazing, you, wait, love, thank, \\ thanks, see, beautiful, heart, so\end{tabular}                     \\ \midrule
\MyEmoji{\smilingfacewithhearts}      & W07, 2019  & \begin{tabular}[c]{@{}l@{}}looking, \textbf{valentines}, tomorrow, \textbf{valentine}, \\ ever, amazing, pretty, sweet, beautiful, you\end{tabular} \\ \bottomrule
\end{tabular}
}
\caption{Highly-associated words (by PMI) of emoji \MyEmoji{\partyingface} and \MyEmoji{\smilingfacewithhearts} in different weeks. The semantics of the associated words coincide with the external events happened at that week.}
\label{tab:top_pmi}
\end{table}

From Table \ref{tab:top_pmi}, we observe that for emoji \MyEmoji{\partyingface} from the 51st week of 2018 to the 1st week of 2019, the words about the New Year event such as ``\textit{happynewyear}'' and ``\textit{2019}'' appear in the associated words, and for emoji \MyEmoji{\smilingfacewithhearts}, the associated words about Valentine's days show in the 7th week of 2019. The observation verifies our hypothesis and suggests that external events can influence the adoption of new emojis. 

\subsection{Influencing Factors of Emoji Diffusion}
The popularity discrepancy of emojis in the same version raises the question of what influence the diffusion process. 
Since emojis can be considered a digital innovation, we apply Rogers' diffusion of innovation theory \cite{rogers2014diffusion}, which describe the diffusion network and innovation itself as the influencing factors. We model each factor and describe their correlation with the new emojis' popularity.

\noindent \textbf{Community size:} \quad
Tracing the diffusion network is extremely challenging, because our tweets is collected through the Twitter's 1\% sampling API, and the low access rate make it impossible to construct the diffusion network. However, previous work suggests that hashtags can indicate community identity in social networks \cite{yang2012whatyoutag, zhou2022emoji}. We thus use the hashtags cooccuring with emojis as the proxy of the early adapter community of the new emojis, and more popular hashtags mean a larger size of communities. 

The Spearman's correlation \cite{hauke2011comparison} of hashtag popularity in the early period and emoji popularity in the late period is 0.580, 0.530, and 0.370 for emojis of Emoji 11.0, 12.0 and 13.0. It indicates that a larger early adopter community may promote the emoji diffusion. (Detailed analysis in Appendix \ref{sec:community_emoji}.) 

\noindent \textbf{Emoji semantics} \quad
Emojis' semantics can be considered the innovation itself, so we hypothesize that emojis with more popular semantics are more adopted\cite{ai2017untangling}). To proxy the popularity of emoji semantics and avoid the circular reasoning, we prompt GPT-4 \cite{openai2023gpt4} to generate words with semantics similar to given emojis and count these words' frequency in the early stage. 

The Pearson correlation \cite{cohen2009pearson} of word occurrence and emoji popularity is 0.413, 0.812 and 0.11 for emojis from Emoji 11.0, 12.0 and 13.0. The significant correlation for Emoji 11.0 and 12.0 supports our hypothesis. For Emoji 13.0, the correlation is weak, likely because users do not directly state the semantics inside the emojis. For example, for \MyEmoji{\pinchedfingers} (pinched fingers), GPT-4 shows us similar words as: \textit{gesture, expressive, Italian, emphasis, and talkative}, which may not be presented in users' tweets.
We present the detailed setup of the experiment in Appendix \ref{sec:emoji_semantics_influence}.

\subsection{Emoji Meaning Evolves during Diffusion}
\label{sec:emoji_sentiment}
With the increasing popularity of the new emojis, one may wonder if their meanings are adapted during the diffusion process. However, understanding the emojis' meaning is hard, as will be discussed in Section \ref{sec:attention_words}. To get a first glimpse of the adaption, we focus on the context where emojis are used, and using the lens of sentiment, because one main functionality of using emojis is to convey sentiments \cite{ai2017untangling}.

\begin{figure}[!h]
     \centering
     \includegraphics[width=\linewidth]{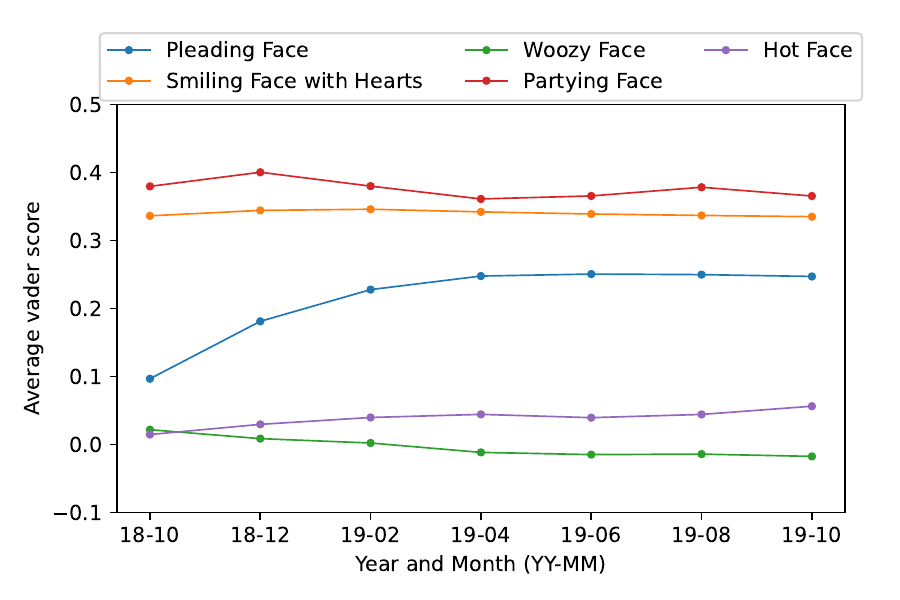}
     \vspace{-2.2em}
     \caption{Average Vader scores of the tweets containing the top 5 popular emojis in Emoji 11.0 from October 2018 to October 2019. 
     }
     \vspace{-0.5em}
     \label{fig:emoji_vader_score}
\end{figure}

To quantify the sentiment context of emojis, 
we calculate the \textit{Vader} score for each tweet \cite{hutto2014vader}, which outputs the sentiment score from -1 (negative) to 1 (positive). The higher the absolute value of the score, the more sentimental the tweet is. We average the Vader score of all tweets with an emoji in two months as the emoji's sentiment score in that time period and visualize its trend of the top 5 popular emojis from Emoji 11.0 every two months in Figure \ref{fig:emoji_vader_score}. 

\begin{figure}[t]
     \centering
     \includegraphics[width=\linewidth]{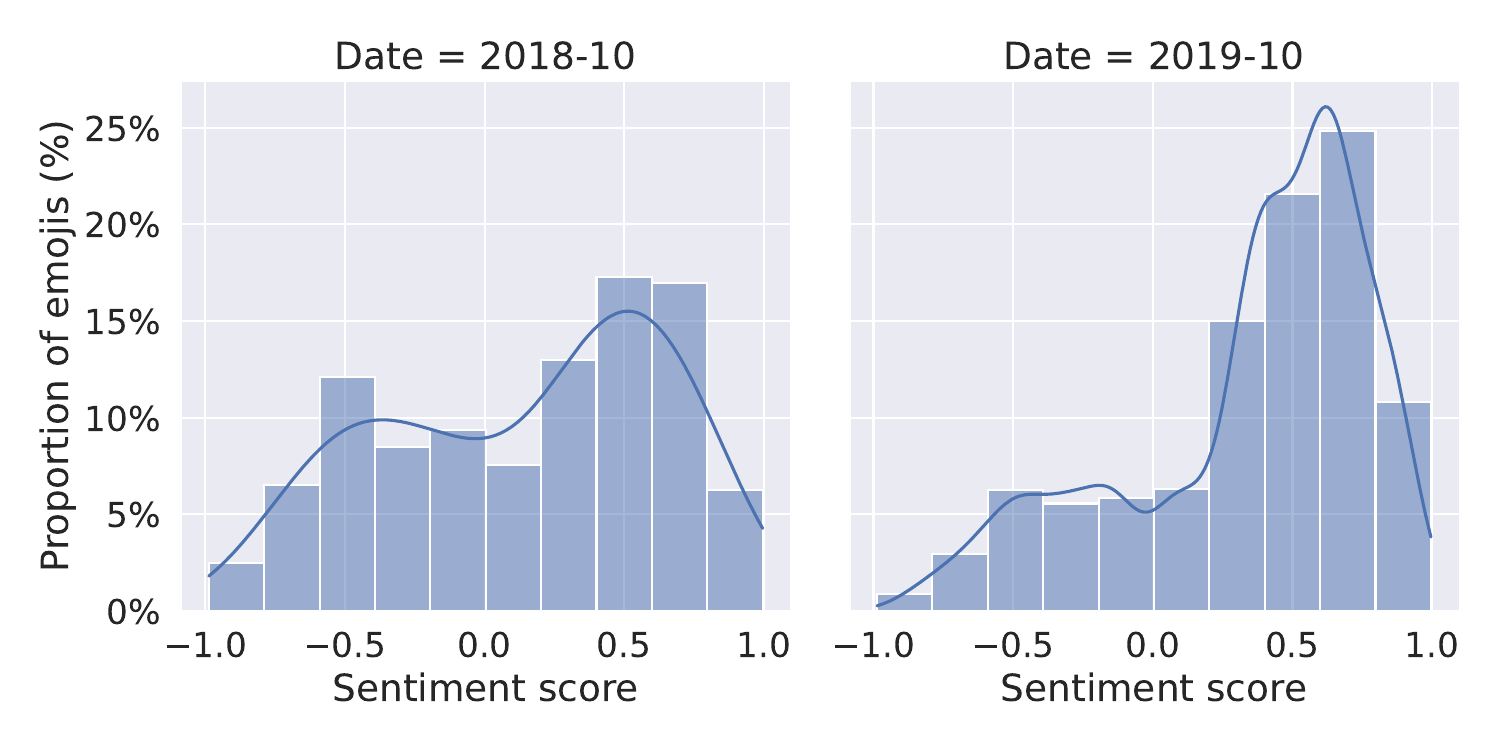}
     \vspace{-2.5em}
     \caption{Distribution of Vader scores of tweets with emoji \MyEmoji{\pleadingface} in October 2018 and October 2019. During one year period, \MyEmoji{\pleadingface} concentrate more on positive tweets.}
     \vspace{-1em}
     \label{fig:pleading_vader_score}
\end{figure}

For most emojis, their sentiment scores remain constant, hinting that the user's understanding of most emojis is unchanged during the adaption. However, for the emoji \MyEmoji{\pleadingface} (pleading face), its sentiment score grew continuously in the first 8 months after its release, suggesting that users use \MyEmoji{\pleadingface} in increasingly positive sentiment context. We further visualize the Vader score distribution of tweets containing \MyEmoji{\pleadingface} in two time periods a year apart (October 2018 and 2019) in Figure \ref{fig:pleading_vader_score}.

Initially (October 2018), many users use the emoji \MyEmoji{\pleadingface} in negative tweets; but during diffusion, more users tend to use it in positive tweets.
We hypothesize that in the early adoption of \MyEmoji{\pleadingface}, users used this emoji to convey sad feelings, but later, the emoji evolved to convey moved or other positive feelings. Indeed, we see that the associated words (measured by PMI) with the emoji \MyEmoji{\pleadingface} changes over time, which provide initial evidence that supports the hypothesis. In October 2018, \textit{sad, sorry, idk} are among the sentimental words with highest PMI, while a year later, we see more positive words such as \textit{prettiest, precious, cutest}. 
(The details are shown in Table \ref{tab:emoji_sentiment_pmi} in Appendix \ref{sec:supplementary}.) Through the lens of sentiment, we observe that the emojis' meaning may evolve during the diffusion process.

\section{Interpret Emojis with Language Models}

The analysis so far shows that the semantic of emojis not only affect their diffusion process, but also evolves during the diffusion process, both of which highlights the importance of an effective framework to interpret new emojis dynamically. 
Although ChatGPT is capable of showing us similar words to emojis \cite{openai2023gpt4, zhou2024emojis}, it lack the understanding based on the application scenario, and cannot capture the semantic evolution during the diffusion. 
Moreover, the pre-training data of ChatGPT may not cover the recently-released emojis.

To address these challenges, we utilize the corpus containing new emojis and open-source language models (LMs) to investigate the application scenario and semantic meaning of emojis. Previous researchers used LMs in interpretation work \cite{lin2023text, zhou2023explore, romanou2023crab} and relied on the attention mechanism to explore the inner association of the fine-tuning data \cite{wang2021identifying}. In this section, we use the attention score method and the cross-dataset inference method to explore words and old emojis with semantics similar to new emojis.

\subsection{Interpretation with High-attention Words}
\label{sec:attention_words}
Attention scores are a well-studied interpretability method to identify important tokens for LMs to make the decision \cite{clark2019does, wang2021identifying}. To understand what words are specifically associated with newly created emojis, we design an emoji prediction task and extract high-attention words to reveal the emoji meaning. 

\subsubsection{Attention Calculation}
Formally, we first construct an emoji classification dataset with the input space $x\in\mathcal{X}$ and the pre-defined emoji label $y$. To better distinguish the word association between new and old emojis, the tweet with label 0 means that the tweet contains the old emoji, and with label $1 \leq k \leq n$ means that the tweet contains the new emoji $k$. Denote $f_e$ as the fine-tuned model in the emoji classification dataset (specifically the Roberta model for our experiments) \cite{liu2019roberta}. For each input tweet $x_i \in \mathcal{X}$ with tokens $\{t^1_i, \cdots, t^m_i\}$, where $m$ is the token number, we adopt $f_e$ in the input $i$ and obtain the attention scores $\{a^1_i, \cdots, a^m_i\}$ and the emoji prediction $f_e(i)$. Since for Roberta model, the embeddings of the [CLS] token in the last layer are used to make the prediction, we compute the scores $a^m_i$ as the average attention scores of the $m^{th}$ token to the [CLS] token across different heads. For the overall attention scores $\overline{a_{kt}}$ of the token $t$ for the emoji $k$, we extract the sentences with prediction $f_e(x_i) = k$ and average the attention scores of the token $t$ in these extracted sentences, which can be formulated as
\begin{equation*}
    \overline{a_{kt}} = \frac{\sum_{x_i\in\mathcal{X}}\mathbbm{1}(f_e(x_i) = k) \cdot \sum_{j=1}^{m} (a^j_i \cdot \mathbbm{1}(t^j_i = t))}{\sum_{x_i\in\mathcal{X}}\mathbbm{1}(f_e(x_i) = k) \cdot \sum_{j=1}^{m} \mathbbm{1}(t^j_i = t)}
\end{equation*}
where $\mathbbm{1}$ is the indicator function.
With the overall attention scores, for each emoji $k$, we can extract the keywords with the highest attention scores $\overline{a_{kt}}$ to understand the semantics of the new emojis.

\subsubsection{Experiment Setup and Results}
To verify the effectiveness of the attention method, we experiment on 6 emojis from Emoji 13.0: \MyEmoji{\smilingfacewithtear} (smiling face with tear), \MyEmoji{\ninja} (ninja), \MyEmoji{\magicwand} (magic wand), \MyEmoji{\pinchedfingers} (pinched fingers), \MyEmoji{\coinemoji} (coin), \MyEmoji{\peoplehugging} (people hugging). We extract the tweets with emojis from April 2022 to May 2022 and construct a balanced dataset with a total of 50,000 tweets and 7 labels (label 0 for old emojis and label 1 to 6 for new emojis). We fine-tune the Roberta model pretrained on tweets \cite{barbieri2020tweeteval} (\texttt{twitter-roberta-base} \footnote{\url{https://huggingface.co/cardiffnlp/twitter-roberta-base}}) on the emoji prediction dataset with the split 8:1:1 and obtain the test accuracy 66.98\%. We present the top 10 words with highest attention scores in the second column of Table \ref{tab:as_emojis} and the words recognized by the authors with similar semantics are highlighted in bold.

\begin{table}[t]
\small
\centering
\resizebox{\columnwidth}{!}{%
\begin{tabular}{cll}
\toprule
Emoji & Top 10 Attention Score Words                                                                                                   & Top 3 Inference Score Emojis \\ \midrule
\MyEmoji{\smilingfacewithtear}      & \begin{tabular}[c]{@{}l@{}}same, its, so, me, \textbf{no}, \\ \textbf{why}, this, \textbf{please}, \textbf{oh}, i\end{tabular}                                & \MyEmoji{\pensiveface} (16.7), \MyEmoji{\pleadingface} (8.5), \MyEmoji{\wearyface} (6.1)                       \\ \midrule
\MyEmoji{\ninja}      & \begin{tabular}[c]{@{}l@{}}days, account, \textbf{ninja}, both, \\ \textbf{assassins}, who, \textbf{samurai},  \\ coin, \textbf{gaming}, website\end{tabular}  &  \MyEmoji{\sparkles} (13.3), \MyEmoji{\eyes} (13.2), \MyEmoji{\smilingfacewithsmilingeyes} (10.2)                       \\ \midrule
\MyEmoji{\magicwand}      & \begin{tabular}[c]{@{}l@{}}\textbf{magic}, \textbf{wish}, \textbf{magician}, \\ \textbf{wizard}, follow, hours, tweet, \\ special, \textbf{light}, recent\end{tabular} &  \MyEmoji{\sparkles} (22.6), \MyEmoji{\collision} (10.1), \MyEmoji{\fire} (9.9)                      \\ \midrule
\MyEmoji{\pinchedfingers}      & \begin{tabular}[c]{@{}l@{}}this, puff, the, that, \textbf{kiss}, \\ another, you, \textbf{art}, \textbf{perfect}, \\ \textbf{please}\end{tabular}             &   \MyEmoji{\hearteyes} (10.2), \MyEmoji{\fire} (7.8), \MyEmoji{\hundredpoints} (6.1)                     \\ \midrule
\MyEmoji{\coinemoji}      & \begin{tabular}[c]{@{}l@{}}start, \textbf{billion}, \textbf{coins}, \textbf{token}, \\ \textbf{coin}, \textbf{money}, proof, gob, \\ hours, follow\end{tabular}        &  \MyEmoji{\collision} (25.1), \MyEmoji{\partypopper} (16.4), \MyEmoji{\fire} (10.5)                        \\ \midrule
\MyEmoji{\peoplehugging}      & \begin{tabular}[c]{@{}l@{}} \textbf{thanks}, \textbf{hugs}, my, \textbf{hug}, \\ \textbf{good}, you, \textbf{hugging}, \textbf{dear}, \\ \textbf{friend}, \textbf{happy}\end{tabular}              &   \MyEmoji{\pleadingface} (11.2), \MyEmoji{\blueheart} (8.9), \MyEmoji{\sparkles} (8.3)                      \\ \bottomrule
\end{tabular}
}
\vspace{-0.7em}
\caption{Words with high attention scores and old emojis with high inference scores for new emojis from Emoji 13.0. High attention scores suggest words with similar semantics as emojis and indicate the application setting of emojis. Emojis with high inference scores represent the old emojis with similar meaning to the new emojis.}
\vspace{-0.5em}
\label{tab:as_emojis}
\end{table}


Table \ref{tab:as_emojis} demonstrates that attention scores are effective in identifying words semantically similar to specific emojis. These words not only mirror the primary meaning of the emojis but also reveal their application scenarios and extended interpretations. For entity-related emojis such as \MyEmoji{\ninja}, \MyEmoji{\magicwand}, and \MyEmoji{\coinemoji}, their high-attention words extend beyond their direct symbolism. For instance, the word ``gaming'' associated with \MyEmoji{\ninja} highlights its frequent use in the context of action video games. Similarly, the term ``token'' linked with \MyEmoji{\coinemoji} (coin) suggests its application in representing Bitcoin tokens, indicating a broader usage beyond its conventional meaning.


Regarding sentiment-related emojis, high-attention score words encapsulate the sentiments embedded in the emojis. For instance, the word ``perfect'' associated with \MyEmoji{\smilingfacewithtear} implies positive sentiment, while ``why'' and ``no'' linked to \MyEmoji{\pinchedfingers} suggests negative sentiments. The qualitative results in Table \ref{tab:as_emojis} demonstrate the effectiveness of using the attention mechanism to probe the application setting and the extensive meaning of the new emojis.

\subsection{Interpret Emojis with Old Emojis}
\label{sec:old_emojis}
The words with high-attention scores can not fully capture the sentiments inherent in emojis. For example, the degree of negative sentiment conveyed by \MyEmoji{\smilingfacewithtear}, or the specific sentiment associated with \MyEmoji{\ninja} and \MyEmoji{\coinemoji}, remains ambiguous. Besides exploring words with similar semantics, we employ the rich and complex sentiments encoded in old emojis to interpret the newly created emojis. 

\begin{figure}[t]
     \centering
     \includegraphics[width=\linewidth]{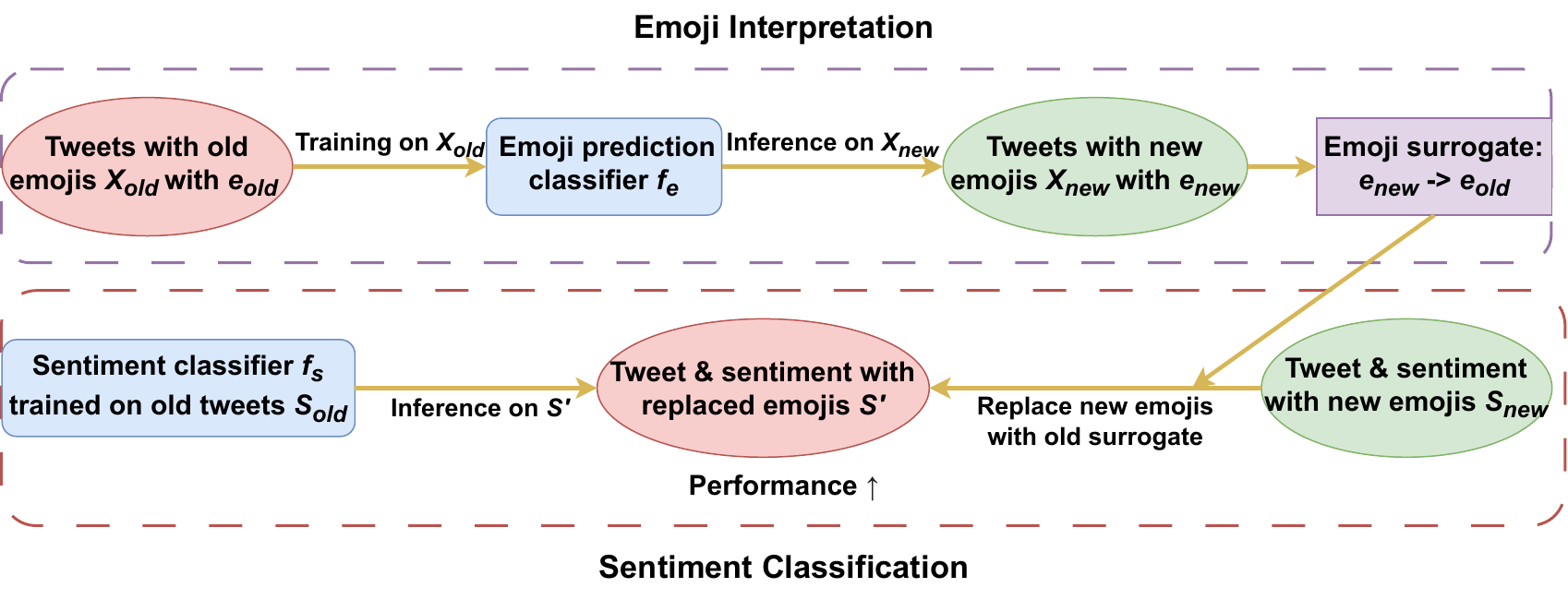}
     \vspace{-1.5em}
     \caption{The upper half shows the framework of using old emojis to interpret new emojis and the lower half presents the pipeline of replacing new emojis in the sentiment classification dataset to old similar emojis to enhance the prediction results.}
     \vspace{-1em}
     \label{fig:emoji_downstream}
\end{figure}

We utilize LMs to conduct cross-version inference to explore old emojis with semantics similar to a new targeted emoji. If the semantics and syntactic features of the text containing two emojis are similar, the semantics of two emojis are also similar. 
Our method is to first fine-tune the LMs on an emoji classification dataset with old emojis as the labels, and we use the fine-tuned LMs to do the inference on tweets containing new emojis. 
If LMs predict tweets with a new emoji to contain another old emoji, it means that two emoji share a similar text context distribution, indicting similar semantics. 
The pipeline of the interpretation framework is shown in the upper half of Figure \ref{fig:emoji_downstream}. 

\subsubsection{Cross-Version Analysis}
We construct an emoji classification dataset with the input tweets $\mathcal{X}_{old}$ and the pre-defined emoji labels from old emojis $\{e^0_{old}, e^1_{old}, \cdots, e^n_{old}\}$. Pre-trained LM $f_e$ is fine-tuned in tweet collection $\mathcal{X}_{old}$. We construct another tweet dataset $\mathcal{X}_{new}$, where each tweet contains the new emojis $\{e^0_{new}, e^1_{new}, \cdots, e^m_{new}\}$ and ask the fine-tuned LM $f$ to do the inference on each tweet $x_{new} \in \mathcal{X}_{new}$. The prediction on the tweet $x_{new}$ containing the emoji $y_{new}$ is: $f_e(x_{new}) = \operatorname*{argmax}_{e_{old}} p(e_{old} | x_{new})$.

The semantic similarity between an old emoji $e_{old}$ and a new emoji $e_{new}$ can then be quantified as the proportion of predictions equal to $e_{old}$:

\begin{equation}
\vspace{-0.5em}
\resizebox{\columnwidth}{!}{%
$
    score(e_{old}, e_{new}) = \frac{\sum_{x_{new}} \mathbbm{1}(y_{new} = e_{new}, f_e(x_{new}) = e_{old})}{\sum_{x_{new}} \mathbbm{1}(y_{new} = e_{new})}.
    \label{eq:simi}
$
}
\end{equation}

We sort the old emojis for each new emoji by the calculated inference $score(\cdot)$ values and for each new emoji, extract three old emojis with the highest inference scores, named as emoji surrogates.

\subsubsection{Experiment Setup and Results}
We first construct $\mathcal{X}_{old}$ with tweets from April 2020 to May 2020, when Emoji 13.0 does not appear on Twitter. 
We extract 10,000 tweets for every emoji in top 32 old emojis from April to May and another 10,000 tweets without any emoji. We form a training dataset $\mathcal{X}_{old}$ with 330,000 tweets and 33 labels (32 emojis + no emoji). The pre-trained Roberta model $f_e$ is fine-tuned in $\mathcal{X}_{old}$ and we use $f_e$ to infer on the test dataset constructed with tweets from 2022 containing the emojis from Emoji 13.0 in Table \ref{tab:as_emojis}. We calculate the inference score between the old and new emojis as Equation \ref{eq:simi} and present the top 3 similar ones for each new emoji, as well as the inference score in the third column in Table \ref{tab:as_emojis}.


Table \ref{tab:as_emojis} reveals that preexisting emojis effectively reflect the semantic content, particularly the sentiment dimension, of new emojis. For sentiment-related emojis such as \MyEmoji{\smilingfacewithtear}, older emojis such as \MyEmoji{\pensiveface} (pensive face), \MyEmoji{\pleadingface} (pleading face) and \MyEmoji{\wearyface} (weary face) encapsulate the blend of sadness and begging inherent in the new emoji. Furthermore, in the case of entity-related emojis, while it may be challenging to pinpoint analogous emojis that precisely capture an entity's characteristics, it is feasible to discern the underlying sentiments these emojis convey. For example, the \MyEmoji{\ninja} (ninja) emoji, in relation to existing emojis such as \MyEmoji{\sparkles} (sparkles), \MyEmoji{\eyes} (eyes), and \MyEmoji{\smilingfacewithsmilingeyes} (smiling face with smiling eyes), reveals the positive sentiment in \MyEmoji{\ninja} in usage.

In the next section, we show how cross-version analysis serves downstream tasks. We find that simply substituting new emojis with old emoji surrogates can significantly increase the accuracy in the sentiment classification task of fine-tuned LMs.

\subsection{Replace Emojis for Sentiment Prediction}
Sentiment classification is a well-studied task and is widely used in deployed systems. Emojis, which encode rich sentiments, have been shown to be important for enhancing the performance in the sentiment classification \cite{chen2019emojipowered}. The current state-of-the-art (SoTA) method for the sentiment classification on tweets is to fine-tune the pre-trained LMs on a training dataset \cite{barbieri2020tweeteval}. However, the newly-created emojis not in both fine-tuning and pre-training data raise a challenge for this method, which may prevent the model from making precise predictions. 

Instead of fine-tuning models on the data with new emojis again, requiring a high computational cost, we propose a method to directly substitute the new emoji with old emoji surrogates without parameter updates. The sentiments encoded in emoji surrogates could compensate for the loss of semantics in the unknown emojis. We present the emoji substitution process in the lower half of Figure \ref{fig:emoji_downstream}.

\subsubsection{Methodology}
Suppose that we have a sentiment classification dataset $\mathcal{S}_{new}$ that contains new emojis and each input has the sentiment label $y$. Given a sentiment classifier $f_s$ fine-tuned in $\mathcal{S}_{old}$, where the new emojis are not in $\mathcal{S}_{old}$, for each tweet $s_i = [t^1_i, t^2_i, \cdots, e_{new}, \cdots, t^m_i] \in \mathcal{S}_{new}$, we replace the new emoji $e_{new}$ in tweet $s_i$ with the old emoji surrogates (top 3 old emojis sorted by the inference scores) and obtain a dataset $\mathcal{S}^{'}$ with $s^{'}_i = [t^1_i, t^2_i, \cdots, e^1_{old}, e^2_{old}, e^3_{old}, \cdots, t^m_i] $. Then we feed the tweet $s^{'}_i$ to the classifier $f_s$ to get the sentiment prediction, which is $f_s(s_i) = \operatorname*{argmax}_{y} p(y | s^{'}_i)$.

\subsubsection{Experiment Setup}

Existing sentiment classification datasets with emojis did not contain the newly created emojis. Therefore, we collect tweets from 2020 and 2022, respectively, as the dataset with old and new emojis.
Due to the superiority of LLMs over human annotators \cite{gilardi2023chatgpt}, we utilize GPT3.5 to annotate the label for tweets \cite{ouyang2022training}. Given a tweet input $s$, and $\mathcal{Y} = \{positive, neutral, negative\}$, ChatGPT is asked to choose a label. We repeat the annotation process twice with temperature 0.7 and keep the examples with labels agreed by two LLM annotators. 

We first randomly collect tweets from April 2020 to May 2020 and then use ChatGPT to label the sentiment of each tweet as $\mathcal{S}_{old}$. We first split $\mathcal{S}_{old}$ into training, validation, and test dataset by 8:1:1 with 35,388 tweets. We fine-tune the Roberta classifier (\texttt{twitter-roberta-base}) on $\mathcal{S}_{old}$ as $f_s$, pre-trained on tweets before August 2019 \cite{barbieri2020tweeteval}. 
The fine-tuned model $f_s$ can achieve an accuracy of 67.52\% in test data from 2020.
We repeat the collection and labeling process in tweets from April 2022 to May 2022 to form $\mathcal{S}_{new}$.

Accuracy of ChatGPT in labeling sentiment can be verified using existing data. We randomly select 1,000 tweets from SemEval-2015 Task 10 \cite{rosenthal2015semeval} and the accuracy of ChatGPT on sentiment prediction is 77.29\%. The average agreement between human annotators and the gold standard annotations in the test dataset is 75.70\%. Therefore, ChatGPT annotations are reliable due to the comparable agreement to the human annotators.

We randomly select 8 emojis from Emoji 13.0, and 4 of them are sentiment-related, and the others are entity-related. We collect the tweets from $\mathcal{S}_{new}$ with the selected emojis. 
We prepare a baseline method to substitute the new emoji with the emoji name (words) to represent the emoji semantics. We present the accuracy on the original tweets, on the emoji replacement tweets, and on the word replacement tweets for each specific emoji in Table \ref{tab:emoji_replacement}. 

\subsubsection{Results}

\begin{table}[t]
\small
\centering
\resizebox{\columnwidth}{!}{%
\begin{tabular}{cccccc}
\hline
\multicolumn{6}{c}{Emoji 13.0 (sentimental)}                                \\ \hline
emoji & surrogates & \# test & ori acc        & replaced acc & word acc   \\
\MyEmoji{\smilingfacewithtear}          &   \MyEmoji{\pensiveface}\MyEmoji{\pleadingface}\MyEmoji{\wearyface}            & 23,080      & 63.43          & \textbf{63.90}  & 59.87 \\
\MyEmoji{\disguisedface}          &     
\MyEmoji{\thinkingface}\MyEmoji{\eyes}\MyEmoji{\pensiveface}
        & 1,908       & 48.69          & 63.87 & \textbf{67.26} \\
\MyEmoji{\peoplehugging}          &     
\MyEmoji{\pleadingface}\MyEmoji{\blueheart}\MyEmoji{\sparkles}
        & 6,580       & 80.32          & \textbf{90.90} & 85.88 \\
\MyEmoji{\pinchedfingers}          &     
\MyEmoji{\hearteyes}\MyEmoji{\fire}\MyEmoji{\hundredpoints}& 6,972       & 79.75          & \textbf{82.80} & 67.76 \\
\hline
\multicolumn{6}{c}{Emoji 13.0 (entity)}                                     \\ \hline
\MyEmoji{\ladder}          &  \MyEmoji{\thumbsup}\MyEmoji{\fire}\MyEmoji{\collision}               & 614         & 22.05 & 18.10     &  \textbf{23.43}   \\
\MyEmoji{\ninja}          &   
\MyEmoji{\sparkles}\MyEmoji{\eyes}\MyEmoji{\smilingfacewithsmilingeyes} & 6,649        & 55.03          & \textbf{56.86} & 54.06 \\
\MyEmoji{\coinemoji}          &   
\MyEmoji{\collision}\MyEmoji{\partypopper}\MyEmoji{\fire}& 3,209        & 52.32          & \textbf{61.92} & 59.58 \\
\MyEmoji{\blueberries}          &      
\MyEmoji{\collision}\MyEmoji{\eyes}\MyEmoji{\sparkles} & 922         & 65.72          & \textbf{67.25} & 68.66 \\ \hline
\end{tabular}
}
\vspace{-0.7em}
\caption{Results of emoji replacement method on the sentiment prediction task for emojis from Emoji 13.0. The fine-tuned models are pre-trained on data without new emojis. \textit{ori acc}, \textit{replaced acc}, and \textit{word acc} represent the accuracy on original test data, replacing new emojis to emoji surrogates, and replacing new emojis with emoji names, respectively. 
}
\vspace{-1em}
\label{tab:emoji_replacement}
\end{table}

From Table \ref{tab:emoji_replacement}, when replacing the new emoji with old emojis with similar semantics, we can observe a notable performance increase for the model $f_s$, with a relative improvement of 4.94\%, 6.92\% on emojis from Emoji 13.0 (sentimental) and Emoji 13.0 (entity), respectively. The relative improvement on the entity-related emojis is more significant, which indicates that the explore old emoji surrogates precisely suggest the sentiment embedded in the entity-related emojis, leading to better results in the sentiment prediction. Compared to replacing the new emoji with the emoji name, our emoji replacement method can outperform in most cases, and the possible reason could be attributed to words from the emoji name, which can not fully characterize the complex sentiments for emojis.  

We repeat the experiments on the Roberta model with new emojis in the pre-trained corpus (\texttt{twitter-roberta-base-2022-154m}, pretrained on tweets before December 2022) \cite{loureiro2023tweet}, and present the results in Table \ref{tab:new_model_emoji_replacement} in Appendix. Compared with models pre-trained on tweets without new emojis, the average improvement of replacing new emojis becomes smaller or even negative, which follows our expectation that the fine-tuned model $f_s$ has learned the semantics of new emojis in the pre-training data and the substitution of emojis causes loss of information. 
\section{Implications}
\label{sec:limitation}



Our research reveals the patterns of emoji diffusion and proposes a framework to understand the semantics of new emojis. 
For model developers, our emoji replacement framework can provide an effective way to increase the generalization of the fine-tuned LLMs on texts containing new emojis without further parameter updates, which can also be extended to other NLP tasks. For future emoji researchers, our work provides a pipeline to explore the pattern of new emoji diffusion and proposes a framework to utilize open source LMs to interpret the emoji semantics and usage scenarios.

Moreover, our work provides an insight of the diffusion patterns on a new language unit in social networks, which can inspire the future researchers to generalize our diffusion exploration pipeline and interpretation framework on other new tokens, such as memes and viral hashtags. Along with our work, we expect that the exploration study on the language unit diffusion can reveal the evolution mechanism of the content in digital communities
\section{Conclusion}
In this work, we explore the pattern of emoji diffusion by studying recently created emojis. We show that the external event can influence the short-period frequency and that some emojis may have an evolution in the sentiments when adopted by more users. 
Then, we propose a framework to use the words or existing emojis to interpret the new emojis. Finally, we show the effectiveness of our framework by replacing new emojis with similar old emojis to improve the sentiment prediction task. The promising results indicate that our framework for interpreting emojis can be used to increase the generalizability of trained models.

\section{Limitations}

There are three limitations in our work. 
First, both our exploration of emoji diffusion and the following semantic analysis regard a type of emoji as a whole at a macro level, but it is not clear how an emoji diffuses by the interaction between users due to the lack of complete network data for Twitter. Considering the current API policy of the Twitter platform, we leave that part to future work by exploring other online communication datasets, such as Slack.

Second, when analyzing the characteristics of new emojis, in this paper, we only focus on the semantic features of emojis, also the focus of the previous work, but we find that the visual features could also be an important feature for emoji diffusion and may also influence the emoji semantics. For example, after one year of diffusion, the number of tweets with \MyEmoji{\whiteheart} (white heart) is 10 times higher than the tweets with \MyEmoji{\brownheart} (brown heart). These visual features play an important role in exploring emoji usage and downstream emoji applications. For future work, it could be interesting to use these visual features, combining the text features to predict the popularity of emojis.

Third, in addition to sentiment classification, the downstream tasks with emojis also include hashtag prediction, gender prediction, and GitHub user dropout prediction \cite{chen2018gender, zhou2023emoji, lu2023team}. Future work could extend our interpretation framework to improve the fine-tuned models on these well-defined tasks without updating the model parameters.



\bibliography{custom}

\appendix

\appendix
\section*{Appendix}

\section{Details of Influencing Factor Analysis}
\label{sec:influencing_factors}

\subsection{Influence of Community Size on Emoji Popularity}
\label{sec:community_emoji}

We first identify the early stage of Emoji 11.0, Emoji 12.0, and Emoji 13.0 as July 2018, September 2019, and November 2020, two months after the first appearance, and set the late stage as the month one year later than the early stage. We count the number of co-occurred hashtags of each emoji in the early stage and the number of emojis in the late stage. We present the top 10 mostly used emojis in the late stage from Emoji 13.0 as well as their top 3 popular co-occurred hashtags in Table \ref{tab:emoji_hashtag} and the count of emojis and hashtags in the early and late stages of emoji diffusion. For the co-occurred hashtags for emojis in Emoji 11.0 and 12.0, we show them in Table \ref{tab:emoji_hashtag_otherversion}.

From Table \ref{tab:emoji_hashtag}, we can observe that there is a pattern in which, for popular emojis in the future, the co-occurred hashtags are also popular in the early adopted stage. Since hashtag number and emoji number are in difference scales in our dataset, we measure the Spearman rank correlation coefficient between hashtag number in the early stage and emoji number in the late stage for the top 10 new emojis from Emoji 11.0, 12.0, and 13.0, respectively \cite{zar2005spearman}.  

The Spearman's correlation of co-occurred hashtag number and emoji number is 0.580, 0.530, and 0.370 ($p-value$ = 0.08, 0.11, 0.29) for the top 10 emojis of Emoji 11.0, 12.0 and 13.0, respectively. It indicates that emojis cooccurring with the popular hashtags tend to be used more in the future, and larger communities promote the emoji diffusion. In addition to the characteristics of early adopters and the diffusion network, we are also interested in the influence of the inner characteristics of emojis on emoji diffusion.

\subsection{Influence of Emoji Semantics on Emoji Popularity}
\label{sec:emoji_semantics_influence}

We expect that extracted words with similar semantics as the emoji should be common on the Twitter platform, so our prompt for querying GPT-4 is composed as follows: \textit{Show me five common single words on Twitter with similar semantics to this emoji: \{emoji\}}, where \textit{\{emoji\}} is the new emoji from Emoji 11.0, 12.0, 13.0, such as \MyEmoji{\smilingfacewithhearts} (smiling face with hearts). We count the average word number in the early stage of new emojis (the same month as in Section \ref{sec:community_emoji}) and the emoji number in the late stage (one year later after the early stage). Since we find that the emoji and word count are on the same scale, we calculate the Spearman's rank correlation coefficient and the Pearson correlation coefficient for the logarithm of these two numbers as the measurement of association \cite{zar2005spearman, cohen2009pearson}. We present the word and emoji number of the top 10 popular emojis from Emoji 13.0 in Table \ref{tab:emoji_word} and for similar words for emojis in Emoji 11.0 and 12.0, we show them in Table \ref{tab:emoji_word_otherversion}.

\begin{table}[!htb]
\small
\centering
\resizebox{\columnwidth}{!}{%
\begin{tabular}{|l|l|l|l|}
\hline
Emoji & Co-occurred hashtag                                                                              & \# hashtag (early) & \# emoji (late) \\ \hline
\MyEmoji{\smilingfacewithtear}      & \begin{tabular}[c]{@{}l@{}}PLTPINKMONDAY, PLTPINKSUNDAY, \\ PLTCyberMonday\end{tabular}          & 79982.2            & 67,103          \\ \hline
\MyEmoji{\pinchedfingers}      & \begin{tabular}[c]{@{}l@{}}ATINYDAY, HeartbreakWeather, \\ NiallHoran\end{tabular}               & 18806.0            & 11,583          \\ \hline
\MyEmoji{\peoplehugging}      & \begin{tabular}[c]{@{}l@{}}EveryVoteCounts, StayHome, \\ StaySafe\end{tabular}                   & 7172.9             & 7,829           \\ \hline
\MyEmoji{\coinemoji}      & \begin{tabular}[c]{@{}l@{}}Bitcoin, gold, \\ DeFi\end{tabular}                                   & 10267.2            & 5,900           \\ \hline
\MyEmoji{\magicwand}      & \begin{tabular}[c]{@{}l@{}}MissguidedCyberTreat, PLTPINKMONDAY, \\ ibelieveinfairys\end{tabular} & 11501.1            & 5,355           \\ \hline
\MyEmoji{\ninja}      & \begin{tabular}[c]{@{}l@{}}syurabaikhati, DerApotheker, \\ wolfpac\end{tabular}                  & 1128.8             & 4,510           \\ \hline
\MyEmoji{\pottedplant}      & \begin{tabular}[c]{@{}l@{}}GreenIndiaChallenge, SupportSmallStreamers, \\ RRRMovie\end{tabular}  & 46272.1            & 3,014           \\ \hline
\MyEmoji{\disguisedface}      & \begin{tabular}[c]{@{}l@{}}Election2020, PLTPINKMONDAY, \\ TREASURE\end{tabular}                 & 83920.8            & 2,497           \\ \hline
\MyEmoji{\flatbread}      & \begin{tabular}[c]{@{}l@{}}HappyThanksgivingEve, Design, \\ Emoji\end{tabular}                   & 866.6              & 2,113           \\ \hline
\MyEmoji{\placard}      & \begin{tabular}[c]{@{}l@{}}MerzmenschPresents, LatentVoices, \\ JukeBox\end{tabular}             & 3343.3             & 1,666           \\ \hline
\end{tabular}
}
\caption{Co-occurred hashtags for top 10 popular emojis in Emoji 13.0 in the late stage. \# hashtag (early) represents the hashtag number in the early stage of emoji diffusion, two months after the first appearance of emojis. \# emoji (late) shows the emoji number in the late stage, one year after the early stage.}
\label{tab:emoji_hashtag}
\end{table}

\begin{table}[!htb]

\begin{subtable}[!htb]{\columnwidth}
\centering
\resizebox{\columnwidth}{!}{%
\begin{tabular}{|l|l|l|l|}
\hline
Emoji & Co-occurred hashtag                                                                  & \# hashtag (early) & \# emoji (late) \\ \hline
\MyEmoji{\pleadingface}      & \begin{tabular}[c]{@{}l@{}}SaveShadowhunters, EXO, \\ BTSARMY,\end{tabular}          & 37122.3            & 101,667         \\ \hline
\MyEmoji{\smilingfacewithhearts}      & \begin{tabular}[c]{@{}l@{}}TeenChoice, \\ SouhilaBenLachhab, Cover\end{tabular}      & 70924.1            & 56,449          \\ \hline
\MyEmoji{\woozyface}      & \begin{tabular}[c]{@{}l@{}}Prediction, Democrats, \\ Obamacare\end{tabular}          & 5121.8             & 29,549          \\ \hline
\MyEmoji{\hotface}      & \begin{tabular}[c]{@{}l@{}}Heatwave, SummerSkinSafety, \\ WorldEmojiDay\end{tabular} & 693.8              & 14,650          \\ \hline
\MyEmoji{\partyingface}      & \begin{tabular}[c]{@{}l@{}}Ethereum, cryptocurrency, \\ eth\end{tabular}             & 22075.3            & 13,070          \\ \hline
\MyEmoji{\coldface}      & \begin{tabular}[c]{@{}l@{}}TRuMP, WorldCup2018, \\ UFC231\end{tabular}               & 11416.0            & 2,137           \\ \hline
\MyEmoji{\teddybear}      & \begin{tabular}[c]{@{}l@{}}Directive, ItsACelebration, \\ TeamNGH\end{tabular}       & 76.1               & 616             \\ \hline
\MyEmoji{\firecracker}      & \begin{tabular}[c]{@{}l@{}}MAGA, CRO, \\ Boxing\end{tabular}                         & 8831.0             & 539             \\ \hline
\MyEmoji{\cupcake}      & \begin{tabular}[c]{@{}l@{}}Cakes, baking, \\ weekendreads\end{tabular}               & 1979.1             & 435             \\ \hline
\MyEmoji{\foot}      & \begin{tabular}[c]{@{}l@{}}TMay, Chequers, \\ UK\end{tabular}                        & 1736.25            & 339             \\ \hline
\end{tabular}
}
\caption{Co-occurred hashtags for emojis in Emoji 11.0}
\label{tab:emoji_hashtag_version11}
\end{subtable}

\begin{subtable}[h]{\columnwidth}
\centering
\resizebox{\columnwidth}{!}{%
\begin{tabular}{|l|l|l|l|}
\hline
Emoji & Co-occurred hashtag                                                                    & \# hashtag (early) & \# emoji (late) \\ \hline
\MyEmoji{\whiteheart}      & \begin{tabular}[c]{@{}l@{}}J9, 3YearsWithCBX, \\ StreamLYTLM\end{tabular}              & 448.50             & 85,981           \\ \hline
\MyEmoji{\personstanding}       & \begin{tabular}[c]{@{}l@{}}Tawan\_V, TeamGalaxy, \\ withGalaxy\end{tabular}            & 144.20             & 13,473           \\ \hline
\MyEmoji{\yawningface}      & \begin{tabular}[c]{@{}l@{}}13ReasonsWhy3, 13ReasonsWhy, \\ DeleteFacebook\end{tabular} & 291.60             & 13,127           \\ \hline
\MyEmoji{\brownheart}       & \begin{tabular}[c]{@{}l@{}}thestripperoficial, \\ camren, natiese\end{tabular}         & 2.0                & 9,321            \\ \hline
\MyEmoji{\pinchinghand}      & \begin{tabular}[c]{@{}l@{}}StarTrekDiscovery, Friends, \\ RossGeller\end{tabular}      & 390.0              & 4,433            \\ \hline
\MyEmoji{\dropofblood}      & \begin{tabular}[c]{@{}l@{}}PeriodEmoji, PeriodStigma, \\ PeriodPoverty\end{tabular}    & 12.0               & 3,648            \\ \hline
\MyEmoji{\ringedplanet}      & \begin{tabular}[c]{@{}l@{}}onlyfans, camgirl, \\ cammodel\end{tabular}                 & 793.0              & 2,729            \\ \hline
\MyEmoji{\whitecane}      & \begin{tabular}[c]{@{}l@{}}AccessATE, InstructionalDesign, \\ CADET\end{tabular}       & 9.70               & 2,603            \\ \hline
\MyEmoji{\purplecircle}      & wolvsden                                                                               & 7.0                & 2,393            \\ \hline
\MyEmoji{\otter}      & -                                                                                      & 0                  & 1,936            \\ \hline
\end{tabular}
}
\caption{Co-occurred hashtags for emojis in Emoji 12.0.}
\label{tab:emoji_hashtag_version12}
\end{subtable}

\caption{Co-occurred hashtags for top 10 popular emojis in Emoji 11.0 and 12.0. \# hashtag (early) represents the hashtag number in the early stage of emoji diffusion, two months after the first appearance of emojis. \# emoji (late) shows the emoji number in the late stage, one year after the early stage.}
\label{tab:emoji_hashtag_otherversion}
\end{table}

\begin{table}[!htb]
\small
\centering
\resizebox{\columnwidth}{!}{%
\begin{tabular}{|l|l|l|l|}
\hline
Emoji & Words with similar semantics                                                            & \# word (early) & \# emoji (late) \\ \hline
\MyEmoji{\smilingfacewithtear}     & \begin{tabular}[c]{@{}l@{}}bittersweet, emotional, touched, \\ relieved, grateful\end{tabular} & 1401.8          & 67,103           \\ \hline
\MyEmoji{\pinchedfingers}      & \begin{tabular}[c]{@{}l@{}}gesture, expressive, Italian, \\ emphasis, talkative\end{tabular}   & 190.8           & 11,583           \\ \hline
\MyEmoji{\peoplehugging}  & \begin{tabular}[c]{@{}l@{}}hug, comfort, support, \\ embrace, togetherness\end{tabular}        & 3714.6          & 7,829            \\ \hline
\MyEmoji{\coinemoji}      & \begin{tabular}[c]{@{}l@{}}money, currency, cash, \\ change, gold\end{tabular}                 & 5999.8          & 5,900            \\ \hline
\MyEmoji{\magicwand}      & \begin{tabular}[c]{@{}l@{}}magic, enchantment, spell, \\ wizardry, mystical\end{tabular}       & 609.6           & 5,355            \\ \hline
\MyEmoji{\ninja}      & \begin{tabular}[c]{@{}l@{}}stealthy, mysterious, skilled, \\ warrior, covert\end{tabular}      & 159.4           & 4,510            \\ \hline
\MyEmoji{\pottedplant}      & \begin{tabular}[c]{@{}l@{}}green, leafy, indoor, \\ botanical, decorative\end{tabular}         & 726.6           & 3,014            \\ \hline
\MyEmoji{\disguisedface}      & \begin{tabular}[c]{@{}l@{}}incognito, hidden, undercover, \\ sneaky, disguised\end{tabular}    & 403.8           & 2,497            \\ \hline
\MyEmoji{\flatbread}      & \begin{tabular}[c]{@{}l@{}}bread, flat, pita, \\ naan, food\end{tabular}                       & 674.0           & 2,113            \\ \hline
\MyEmoji{\placard}      & \begin{tabular}[c]{@{}l@{}}sign, protest, message, \\ board, banner\end{tabular}               & 1311.0          & 1,666            \\ \hline
\end{tabular}
}
\caption{Words with similar semantics (from ChatGPT) for top 10 popular emojis in the late stage in Emoji 13.0. \# word (early) represents the word number in the early stage of emoji diffusion, two months after the first appearance of emojis. \# emoji (late) shows the emoji number in the late stage, one year after the early stage.}
\label{tab:emoji_word}
\end{table}

\begin{table}[!htb]
\begin{subtable}[h]{\columnwidth}
\centering
\resizebox{\columnwidth}{!}{%
\begin{tabular}{|l|l|l|l|}
\hline
Emoji & Words with similar semantics                                                              & \# word (early) & \# emoji (late) \\ \hline
\MyEmoji{\pleadingface}      & \begin{tabular}[c]{@{}l@{}}please, sorry, help, \\ sad, desperate\end{tabular}            & 12973.8         & 101,667          \\ \hline
\MyEmoji{\smilingfacewithhearts}      & \begin{tabular}[c]{@{}l@{}}love, happy, adorable, \\ blissful, sweet\end{tabular}         & 33005.4         & 56,449           \\ \hline
\MyEmoji{\woozyface}      & \begin{tabular}[c]{@{}l@{}}dizzy, confused, woozy, \\ drunk, lightheaded\end{tabular}     & 1242.0          & 29,549           \\ \hline
\MyEmoji{\hotface}      & \begin{tabular}[c]{@{}l@{}}hot, sweaty, exhausted, \\ overwhelmed, burning\end{tabular}   & 2578.2          & 14,650           \\ \hline
\MyEmoji{\partyingface}      & \begin{tabular}[c]{@{}l@{}}celebrating, party, woohoo, \\ ecstatic, jubilant\end{tabular} & 1470.0          & 13,070           \\ \hline
\MyEmoji{\coldface}      & \begin{tabular}[c]{@{}l@{}}cold, freezing, shivering, \\ frosty, icy\end{tabular}         & 591.6           & 2,137            \\ \hline
\MyEmoji{\teddybear}      & \begin{tabular}[c]{@{}l@{}}cute, soft, cuddly, \\ plush, adorable\end{tabular}            & 4112.6          & 616             \\ \hline
\MyEmoji{\firecracker}      & \begin{tabular}[c]{@{}l@{}}loud, explosive, bang, \\ pop, fireworks\end{tabular}          & 1273.2          & 539             \\ \hline
\MyEmoji{\cupcake}      & \begin{tabular}[c]{@{}l@{}}sweet, delicious, cute, \\ frosting, treat\end{tabular}        & 5303.2          & 435             \\ \hline
\MyEmoji{\foot}      & \begin{tabular}[c]{@{}l@{}}step, walk, run, \\ sole, toe\end{tabular}                     & 2263.0          & 339             \\ \hline
\end{tabular}
}
\caption{Words with similar semantics for emojis in Emoji 11.0.}
\label{tab:emoji_word_version11}
\end{subtable}

\begin{subtable}[!htb]{\columnwidth}
\centering
\resizebox{\columnwidth}{!}{%
\begin{tabular}{|l|l|l|l|}
\hline
Emoji & Words with similar semantics                                                                   & \# word (early) & \# emoji (late) \\ \hline
\MyEmoji{\whiteheart}      & \begin{tabular}[c]{@{}l@{}}pure, love, clear, \\ sincere, peace\end{tabular}            & 21063.2         & 85,981          \\ \hline
\MyEmoji{\personstanding}      & \begin{tabular}[c]{@{}l@{}}stand, upright, wait, \\ solo, idle\end{tabular}             & 4793.4          & 13,473          \\ \hline
\MyEmoji{\yawningface}      & \begin{tabular}[c]{@{}l@{}}tired, sleepy, bored, \\ exhausted, drowsy\end{tabular}      & 1676.2          & 13,127          \\ \hline
\MyEmoji{\brownheart}      & \begin{tabular}[c]{@{}l@{}}warmth, earthy, comfort, \\ stable, rich\end{tabular}        & 496.6           & 9,321           \\ \hline
\MyEmoji{\pinchinghand}      & \begin{tabular}[c]{@{}l@{}}small, little, slight, \\ precise, minimal\end{tabular}      & 3843.8          & 4,433           \\ \hline
\MyEmoji{\dropofblood}      & \begin{tabular}[c]{@{}l@{}}blood, bleed, drip, \\ red, donate\end{tabular}              & 1376.2          & 3,648           \\ \hline
\MyEmoji{\ringedplanet}      & \begin{tabular}[c]{@{}l@{}}saturn, space, cosmic, \\ orbit, celestial\end{tabular}      & 398.6           & 2,729           \\ \hline
\MyEmoji{\whitecane}      & \begin{tabular}[c]{@{}l@{}}blind, aid, navigate, \\ mobility, independence\end{tabular} & 266.2           & 2,603           \\ \hline
\MyEmoji{\purplecircle}      & \begin{tabular}[c]{@{}l@{}}purple, geometric, round, \\ violet, circle\end{tabular}     & 842.6           & 2,393           \\ \hline
\MyEmoji{\otter}      & \begin{tabular}[c]{@{}l@{}}otter, playful, aquatic, \\ furry, adorable\end{tabular}     & 487.4           & 1,936           \\ \hline
\end{tabular}
}
\caption{Words with similar semantics for emojis in Emoji 12.0.}
\label{tab:emoji_word_version13}
\end{subtable}
\caption{Words with similar semantics (from ChatGPT) for top 10 popular emojis in Emoji 11.0 and 12.0. \# word (early) represents the word number in the early stage of emoji diffusion, two months after the first appearance of emojis. \# emoji (late) shows the emoji number in the late stage, one year after the early stage.}
\label{tab:emoji_word_otherversion}
\end{table}

\subsection{Supplementary Results and Prompt Details}
\label{sec:supplementary}

\begin{table}[!htb]
\centering
\resizebox{\columnwidth}{!}{%
\begin{tabular}{l|l|l}
\hline
Time Period & Top 10 PMI Sentimental Words     & Score Avg.                                                                                                                  \\ \hline
2018-10      & \begin{tabular}[c]{@{}l@{}}\colorbox{blue!47.7}{\strut cry} \colorbox{blue!47.7}{\strut sad} \colorbox{red!31.8}{\strut please} \colorbox{blue!15.3}{\strut miss} \colorbox{blue!57.2}{\strut hate} \\ \colorbox{blue!7.7}{\strut sorry} \colorbox{blue!10.3}{\strut idk} \colorbox{red!40.2}{\strut wish} \colorbox{blue!54.2}{\strut bad} \colorbox{blue!29.6}{\strut stop}\end{tabular}  &    -0.198          \\ \hline\hline
2019-10      &  \begin{tabular}[c]{@{}l@{}}\colorbox{blue!38.18}{\strut sobbing} \colorbox{red!38.2}{\strut protect} \colorbox{blue!47.7}{\strut cry} \colorbox{red!57.2}{\strut prettiest} \colorbox{red!31.8}{\strut pls} \\ \colorbox{red!57.2}{\strut precious} \colorbox{red!47.7}{\strut hug} \colorbox{red!58.6}{\strut cutest} \colorbox{blue!47.7}{\strut sad} \colorbox{red!63.7}{\strut heart}\end{tabular}  &            0.221          \\ \hline
\end{tabular}
}
\caption{Top 10 associated words with positive or negative sentiments of emoji \MyEmoji{\pleadingface} in October 2018 and October 2019. Red and blue highlight positive and negative words, respectively. The darker the background color, the more sentimental the words.}
\label{tab:emoji_sentiment_pmi}
\end{table}

\begin{table}[!htb]
\small
\centering
\resizebox{\columnwidth}{!}{%
\begin{tabular}{ccccc}
\hline
\multicolumn{5}{c}{Emoji 13.x (sentimental)}                                \\ \hline
emoji & surrogates & \# test & ori acc        & replaced acc   \\
\MyEmoji{\smilingfacewithtear}          &   \MyEmoji{\pensiveface}\MyEmoji{\pleadingface}\MyEmoji{\wearyface}            & 23,080      & \textbf{66.57}          & 60.58 \\
\MyEmoji{\disguisedface}          &     
\MyEmoji{\thinkingface}\MyEmoji{\eyes}\MyEmoji{\pensiveface}
        & 1,908       & 57.08          & \textbf{66.18} \\
\MyEmoji{\peoplehugging}          &     
\MyEmoji{\pleadingface}\MyEmoji{\blueheart}\MyEmoji{\sparkles}
        & 6,580       & 86.99          & \textbf{91.57} \\
\MyEmoji{\pinchedfingers}          &     
\MyEmoji{\hearteyes}\MyEmoji{\fire}\MyEmoji{\hundredpoints}& 6,972       & 82.77          & \textbf{83.79} \\
\MyEmoji{\facewithspiraleyes}          &  \MyEmoji{\flushedface}\MyEmoji{\sparkles}\MyEmoji{\horns}               & 6,206         & \textbf{56.86} & 55.15   \\
\MyEmoji{\mendingheart}          &   
\MyEmoji{\sparkles}\MyEmoji{\pleadingface}\MyEmoji{\pensiveface} & 6,160        & 84.10          & \textbf{88.44} \\
\MyEmoji{\faceexhaling}          &   
\MyEmoji{\horns}\MyEmoji{\droplets}\MyEmoji{\woozyface}& 6,390        & \textbf{51.75}          & 40.94 \\
\MyEmoji{\faceinclouds}          &      
\MyEmoji{\sparkles}\MyEmoji{\horns}\MyEmoji{\eyes} & 1,808         & 48.12          & \textbf{56.19} \\ \hline
\multicolumn{5}{c}{Emoji 13.x (entity)}                                     \\ \hline
\MyEmoji{\ladder}          &  \MyEmoji{\thumbsup}\MyEmoji{\fire}\MyEmoji{\collision}               & 614         & \textbf{22.39} & 17.57  \\
\MyEmoji{\ninja}          &   
\MyEmoji{\sparkles}\MyEmoji{\eyes}\MyEmoji{\smilingfacewithsmilingeyes} & 6,649        & 58.02          & \textbf{59.82} \\
\MyEmoji{\coinemoji}          &   
\MyEmoji{\collision}\MyEmoji{\partypopper}\MyEmoji{\fire}& 3,209        & 63.78          & \textbf{66.84} \\
\MyEmoji{\blueberries}          &      
\MyEmoji{\collision}\MyEmoji{\eyes}\MyEmoji{\sparkles} & 922         & \textbf{69.95}          & 69.31 \\ \hline
\end{tabular}
}
\caption{Results of emoji replacement method on the sentiment prediction task for selected emojis from Emoji 13.0 and 13.1 on models pre-trained on data before December 2022, including new emojis in the pre-training data.}
\label{tab:new_model_emoji_replacement}
\end{table}

\end{document}